\documentclass[runningheads]{llncs}
\usepackage[T1]{fontenc}
\usepackage{multirow}
\usepackage{graphicx}
\begin{document}
\title{The impact of using voxel-level segmentation metrics on evaluating multifocal prostate cancer localisation}
%

\author{ Wen Yan$^{1,2}$ 
\and Qianye Yang $^1$
\and Tom Syer$^3$
\and Zhe Min$^1$
\and Shonit Punwani$^3$
\and Mark Emberton$^4$
\and Dean C. Barratt$^1$
\and Bernard Chiu$^2$
\and Yipeng Hu$^1$
        }
        
\authorrunning{}
%
\institute{Centre for Medical Image Computing; Wellcome/EPSRC Centre for Interventional \& Surgical Sciences; Department of Medical Physics \& Biomedical Engineering, University College London, London, U.K. 
\and 
Department of Electrical Engineering, City University of Hong Kong, HKSAR, China.  
\and 
Centre for Medical Imaging, Division of Medicine, University College London, London, U.K.
\and
Division of Surgery  \&  Interventional Science, University College London, London, U.K.}
\maketitle              

\begin{abstract}
Dice similarity coefficient (DSC) and Hausdorff distance (HD) are widely used for evaluating medical image segmentation. They have also been criticised, when reported alone, for their unclear or even misleading clinical interpretation. DSCs may also differ substantially from HDs, due to boundary smoothness or multiple regions of interest (ROIs) within a subject. More importantly, either metric can also have a nonlinear, non-monotonic relationship with outcomes based on Type 1 and 2 errors, designed for specific clinical decisions that use the resulting segmentation. Whilst cases causing disagreement between these metrics are not difficult to postulate, one might argue that they may not necessarily be substantiated in real-world segmentation applications, as a majority of ROIs and their predictions often do not manifest themselves in extremely irregular shapes or locations that are prone to such inconsistency. This work first proposes a new asymmetric detection metric, adapting those used in object detection, for planning prostate cancer procedures. The lesion-level metrics is then compared with the voxel-level DSC and HD, whereas a 3D UNet is used for segmenting lesions from multiparametric MR (mpMR) images. Based on experimental results using 877 sets of mpMR images, we report pairwise agreement and correlation 1) between DSC and HD, and 2) between voxel-level DSC and recall-controlled precision at lesion-level, with Cohen's $\kappa \in [0.49, 0.61] $ and Pearson's $r \in [0.66, 0.76]$ (\textit{p-values}$<$0.001) at varying cut-offs. However, the differences in false-positives and false-negatives, between the actual errors and the perceived counterparts if DSC is used, can be as high as 152 and 154, respectively, out of the 357 test set lesions. We therefore carefully conclude that, despite of the significant correlations, voxel-level metrics such as DSC can misrepresent lesion-level detection accuracy for evaluating localisation of multifocal prostate cancer and should be interpreted with caution.

\end{abstract}

\section{Introduction}
\label{sec:intro}

Prostate cancer is one of the most frequently occurring malignancies in adult men~\cite{sung2021global}. Transrectal ultrasound-guided (TRUS) biopsy is the gold standard for detecting and grading prostate cancer; however, it has side effects such as bleeding, pain, and infection. Multiparametric MR (mpMR) scan has been suggested as a non-invasive assessment tool before a biopsy~\cite{ahmed2017diagnostic}. In addition, mpMR scans can be performed after a biopsy or a treatment to investigate if any cancer found in the prostate has progressed.
Radiologists read mpMR images and give lesions a score of 1-5 by using a Likert scoring system ~\cite{villers2006dynamic}\cite{jung2004prostate} or the Prostate Imaging Reporting and Data System (PI-RADS)~\cite{weinreb2016pi}\cite{dickinson2013scoring}.
Both scoring systems are used to describe the level of evidence for detecting lesions and to standardize radiological assessment.


Recently, deep learning approaches have been proposed for diagnosing patients with suspicion of clinically significant prostate cancer and segment prostate lesions directly from mpMR images~\cite{chIoU2020harnessing}\cite{hambarde2020prostate}\cite{schelb2021comparison}\cite{cao2019prostate}\cite{winkel2020autonomous}. In addition to evaluating patient-level cancer detection - an image classification problem, results from image segmentation using UNet and its variant \cite{ronneberger2015u} have also been reported, as the mpMR-detected lesions may need histo-pathological examination or treatment, for example, through targeted biopsy \cite{kasivisvanathan2018mri}\cite{moore2013standards} 
and focal ablation \cite{ahmed2012focal}\cite{van2018focal}. 
Further examples of mpMR-based procedure planning are discussed in Sec.\ref{sec:application}.

\begin{figure}
     \centering
     \includegraphics[width=0.8\textwidth]{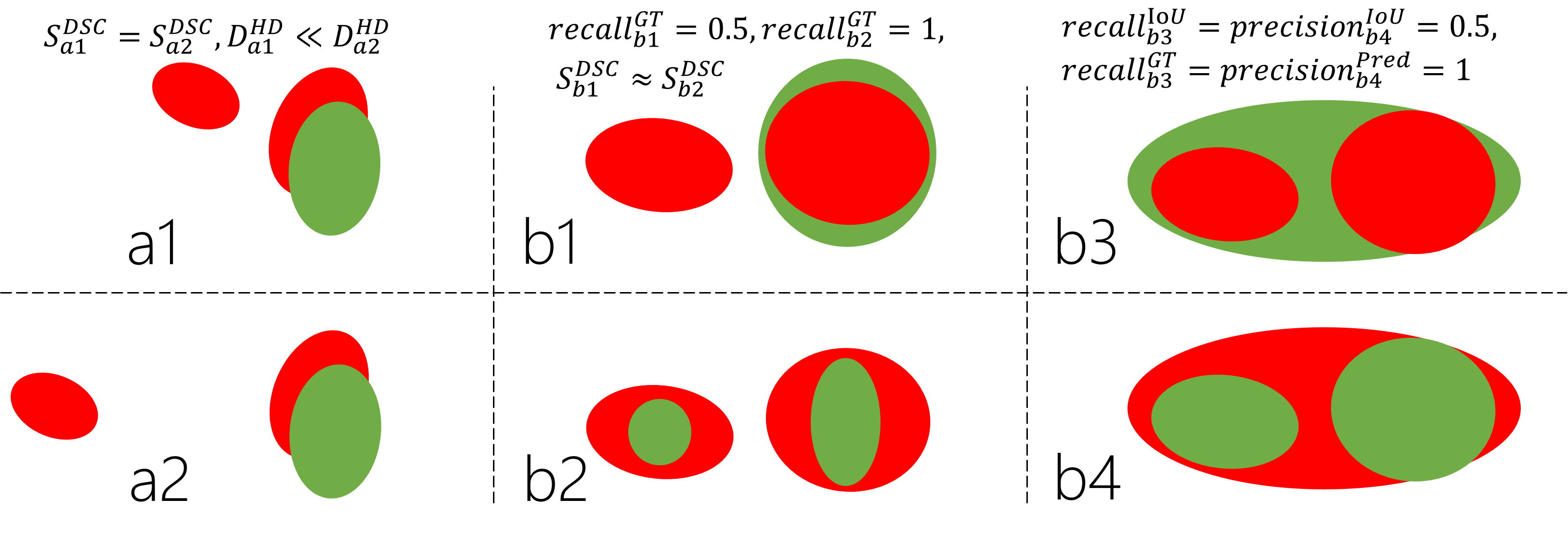}
     \caption{Illustration of the prediction (green) overlaid with the ground-truth lesion (red), 
     details are described in Sec.\ref{sec:method}.}
     \label{fig:my_label}
 \end{figure}

Voxel-level segmentation metrics including Dice similarity coefficient (DSC) and Hausdorff distance (HD), defined in Sec.\ref{sec:metrics_vox_lesion}, have long been established to evaluate segmentation accuracy and reported for this application~\cite{chIoU2020harnessing}\cite{hambarde2020prostate}\cite{schelb2021comparison}\cite{cao2019prostate}\cite{winkel2020autonomous} \cite{menze2014multimodal}.
However, they have also been debated for their consistency and clinical relevance. These limitations are particularly evident with multiple regions-of-interest (ROIs). As illustrated in a1 and a2, Fig.\ref{fig:my_label}, when the number of predicted (green) and ground-truth (red) ROIs differ, the HDs become dependent on the distance from the missed ground-truth to the others, but the DSCs do not. Other factors include boundary smoothness that are more sensitive to HDs than DSCs.

Perhaps more relevant to clinical practice, either DSC or HD can disagree with other metrics that are related to clinical decision making. Examples include lesion-level detection accuracy that is of interest in this study. Shown in Fig.\ref{fig:my_label}, one ground-truth ROI is entirely missed in b1 while both ground-truth ROIs are, albeit partially, detected in b2. These two cases measure to similar DSCs (here, $\approx0.5$). The lesion-level accuracy is important in many biomedical imaging applications. In our applications, described in Sec.\ref{sec:application}, the heterogeneity within individual foci has motivated diagnostic sampling and treating these lesions individually, rather than a uniform coverage of all cancerous regions, partly due to the natural history of the pathology~\cite{van2018focal}\cite{ahmed2012focal}.  

Lesion-level metrics used in object detection literature are arguably appropriate for quantifying lesion detection, even when segmentation algorithms are used, which have produced promising results in this application, perhaps due to well-tuned algorithms such as UNet~\cite{schelb2019classification}. The object detection metrics are discussed in Sec.\ref{sec:metrics_vox_lesion}, together with their limitations in evaluating segmentation output. We propose a new asymmetric lesion-level measure in Sec.\ref{sec:metrics_asymmetric}.

We investigate the following specific questions in this application. \textbf{(1) Do cases, such as those in Fig.\ref{fig:my_label} that cause disagreement between these metrics, exist in clinical data with a competent segmentation network?} and \textbf{(2) To what extent, the presence of these cases affects the ability of voxel-level segmentation metrics in evaluating lesion-level detection?} The correlation between these metrics can be dependent on the radiological/histopathological ground-truth, the adopted segmentation networks and the clinical tasks that utilise these detected lesions, while this study is intended to answer these questions in specific clinical applications described in Sec.\ref{sec:application}. In Sec.\ref{sec:results}, we report both voxel- and lesion-level results from prostate lesion segmentation on clinical mpMR and summarise the overall disagreement levels and correlations at a range of cut-off values. These metric results are presented together with the quantified impact, due to adopting the segmentation metrics, on false-positive and false-negative numbers in this application.

\section{Materials and methods}
\label{sec:method}
\subsection{Prostate lesion segmentation for procedure planning}
\label{sec:application}
The mpMR images were acquired from 850 prostate cancer patients, with a total of 877 studies from [\textit{anonymous trial details}].
All trial patients gave written consents and the ethics was approved as part of the respective trial protocols. Radiologist contours were obtained for all lesions with Likert-scores$\geq$3 and served as ground-truth labels in this study. 192, 325, 232 and 106 studies have 1, 2, 3 and $\geq$4 lesions, respectively. 
Image volumes were resampled to $0.5\times{0.5}\times{1.0} mm^3$ with normalised intensities [0,1], before being centre-cropped to a size of $192\times{192}\times{96}$ voxels using gland segmentation masks, primarily for computational consideration. No other pre-processing was applied.

Three MR volumes, T2-weighted, ADC and diffusion with high-b values, were channel-wise concatenated as the input to a 3D UNet. 877$\times$3 mpMR images were divided into 503, 190 and 184, as training-, validation- and test sets without same patients in different sets. The network was trained with equally-weighted cross-entropy and Dice losses \cite{ma2021loss} for 100 epochs, using an Adam optimiser with a weight decay of 0.0001. The networks and training strategies were otherwise non-optimised for this application to provide a reference performance. While clustering ROIs from voxel-level segmentation algorithms remain an open research question, individual lesions were separated, for the purpose of this study, by testing the neighbouring 26-connected voxels, before filtering out any isolated ROIs with fewer than 8 voxels. 

For targeted biopsy planning, 3-6 biopsy needle positions are planned for individual detected lesions \cite{kasivisvanathan2018mri}\cite{moore2013standards}. Missing clinically-significant lesions could mean missed opportunity for early detection of tumours that are still amenable for less radical treatment. In a number of treatments such as focal therapy \cite{ahmed2012focal}
and nerve-sparing surgery \cite{catalona1990nerve}\cite{rob2010nerve}, accurately identifying individual lesions does not only ensure adequate coverage of cancers, but also supports different function-preserving surgical options, as false-positive lesions could invalidate less-invasive treatment options with less-complications and quicker recovery. In this work, we focus on lesion-level accuracy as an example type of clinically-relevant metrics to examine voxel-level DSC and HD.

\subsection{Voxel-level segmentation metrics}
\label{sec:metrics_vox_lesion}
DSC measures the overlap between the predicted segmentation \(Y_p\) and the ground-truth \(Y_g\), $\mathcal{S}^{DSC}=2\times({Y_p\cap{Y_g}})/{Y_p\cup{Y_g}}$. HD measures the greatest surface distance between the boundaries of the predicted segmentation and the ground-truth. In Sec.\ref{sec:results}, we report the $95^{th}$ percentile of surface distances, denoted as $\mathcal{D}^{HD}$, as a robust alternative. 


\subsection{Lesion-level object detection metrics}
In object detection literature, the accuracy metrics are computed typically for comparing overall performance from different algorithms, by first defining the individual \textit{true-positives} that represent correctly detected ROIs (i.e. instances or, here, lesions), based on an overlap measure intersection over union (IoU), $\mathcal{S}^{IoU}=({Y_p\cap{Y_g}})/{Y_p\cup{Y_g}}$, between the predicted $Y_p$ and the ground-truth $Y_g$ ROIs. The predicted ROIs with greater and less than a given overlap \textit{threshold}  $s^{IoU}$ are considered true-positives and false-positives, respectively. Ground-truth lesions that are not detected by any predicted ROIs are false-negatives. Individual predicted or ground-truth ROIs are counted once with their highest-overlapping counterparts~\cite{electronics10030279}. True-negative ROIs are not defined. These metrics are customarily defined between bounding-boxes representing instances in object detection algorithms, whilst these are reported at voxel-level in Sec.\ref{sec:results} with respect to the evaluated segmentation algorithms.

The lesion-level accuracy can, in turn, be summarised by counting the numbers of true-positive, false-positive and false-negative ROIs, denoted as TP, FP and FN, respectively,
$
precision^{IoU}=TP^{IoU}/(TP^{IoU}+FP^{IoU})
$
and
$
recall^{IoU}=TP^{IoU}/(TP^{IoU}+FN^{IoU})
$
where the superscripts $IoU$ indicate the IoU-based definitions.

Marginalising over varying prediction probability \textit{cut-off} values obtains summary metrics such as average precision, area under the precision-versus-recall curve. However these metrics are not discussed further in this work, as they may be appropriate for comparing different methods but lacks direct clinical interpretation in specific applications \cite{halligan2015disadvantages}. We instead compare precision with controlled recall at individual cut-off values, whereas this cut-off applies on voxel class probabilities in the segmentation task, rather than on ROI (instance) probabilities in a typical object detection algorithm. 

\textbf{Limitations in evaluating multifocal cancer localisation}
Direct applying the object detection metrics was found challenging to interpret in our multifocal cancer application, which considers multiple ROIs of the same type\footnote{This work uses binary segmentation as an example, though the discussion may generalise to multiclass segmentation by considering lesions of different grades separately.} described in Sect\ref{sec:application}. Object detection algorithms allow flexible and many more ROI candidate proposals before thresholding on overlaps. This is in contrast to typical segmentation results, in which no overlapping ROIs are predicted. To balance the flexibility in region proposals and avoiding over-predicting, $TP^{IoU}$, $FP^{IoU}$ and $FN^{IoU}$ are designed with an arguably more stringent criterion. As described above, a) a symmetric overlap measure IoU is used, and b) individual predicted and ground-truth ROIs are not allowed to be counted more than once. For example, one of the two ground-truth ROIs is considered as false-negative in Fig.\ref{fig:my_label}.b3, while the left green ROI is likely to be a false-positive in Fig.\ref{fig:my_label}.b4. We observed that disregarding ROIs with substantial overlap in these two cases, by using object detection metrics on segmentation output, makes it difficult to account for varying detection and coverage levels. In other words, changing cut-off does not differentiate cases, in which both ground-truth ROIs were ``detected" or both predicted ROIs collectively ``covered" the disease area, from others. 

In general, it is its symmetric nature of the overlap measure $\mathcal{S}^{IoU}$ (and $\mathcal{S}^{DSC}$) that lead to $precision^{IoU}$ and $recall^{IoU}$ being insensitive to the different combinations of false-negative- and false-positive \textit{voxels}, where the cut-off applies for segmentation output. As we show in Table~\ref{tab:summary} in Sect.~\ref{sec:results}, different values in $precision^{IoU}$ indeed lead to similar or even the same $recall^{IoU}$ values, vice versa. This indifference to the ``costs'' associated with respective false-negatives and false-positives could have direct clinical consequences. However, adapting the definition of individual ROIs, such that they are amenable to these metrics, is not trivial. 

Furthermore, for the purpose of assessing other metrics, voxel-level segmentation metrics in this work, such insensitive lesion-level metrics may over-estimate their correlation due to limited numerical and statistical precision determined by practical factors such as noise in the data and size of the subjects. 

\subsection{Lesion detection metrics for multifocal segmentation output}
\label{sec:metrics_asymmetric}
To address some of the above-discussed limitations with $precision^{IoU}$ and $recall^{IoU}$ for assessing the segmentation output in this clinical applications, we propose an asymmetric measure, such that the voxel-level cut-off becomes sensitive to, thus practically useful for, balancing the two error types.

For each of $N$ ground-truth lesions $\{Y_g^n\}_{n=1,...,N}$, it is considered as a true-positive lesion if it has overlap with any of the $M$ predictions \{$Y_p^m\}_{m=1,...,M}$, single or multiple, that is greater than a pre-defined overlap threshold $s^{GT}$, otherwise false-negative. Thus, 
$
\mathcal{S}^{GT}=\sum_{m=1}^M (Y_p^m\cap{Y_g^n})/{Y_g^n}
$
where the superscripts $GT$ indicating the ground-truth-based definitions, with which false-positive lesions is not defined. The recall thus can be computed, 
$
recall^{GT}=TP^{GT}/(TP^{GT}+FN^{GT})
$
where $TP^{GT}$ and $FN^{GT}$ are the numbers of true-positive and false-negative lesions using the ground-truth-based definitions, respectively.

For individual predicted lesions $Y_p^m$, a true-positive lesion requires the overlap with ground-truth regions $Y_g^n$ to be greater than $s^{Pred}$, otherwise false-positive. Thus, 
$
\mathcal{S}^{Pred}=\sum_{n=1}^N (Y_g^n\cap{Y_p^m})/{Y_p^m}
$
where the superscripts $Pred$ for the prediction-based definitions and undefined $FN^{Pred}$. Therefore, 
$
precision^{Pred}=TP^{Pred}/(TP^{Pred}+FP^{Pred}).
$

\subsection{Correlation, pairwise agreement and impact on evaluation}
Pearson's $r$ is reported to measure the linear correlation between $\mathcal{S}^{DSC}$ and $\mathcal{D}^{HD}$ and that between voxel-level $\mathcal{S}^{DSC}$ and lesion-level $precision^{Pred}$/$precision^{IoU}$, on 100 boostrapping samples with a sample size of 20, from the holdout set. Cohen's Kappa coefficient $\kappa$ is computed, between two metrics, to measure the level of pairwise agreement on judging the better one from two randomly sampled holdout cases. For example, a higher $\mathcal{S}^{DSC}$ agrees (true) with a higher $precision^{Pred}$ but disagrees (false) with a higher $\mathcal{D}^{HD}$, if measured from the same case of the two. For comparison purposes, an overlap threshold of 0.3 was used for $S^{IOU}$, $S^{Pred}$ and $S^{GT}$, approximating the mean $\mathcal{S}^{DSC}$ on holdout set, with varying cut-off values. Other non-extreme threshold values did not alter the conclusions summarised in Sec.\ref{sec:results}.

The DSC results are used as an example to estimate the false-positive and false-negative cases, using the linearly fitted correlation models (see Fig.~\ref{fig:dicevsPR} for examples). The differences, between these Dice-estimated and the actual errors, provide quantitative evidence of the clinical impact on evaluating lesion localisation using segmentation metrics.  

\begin{table*}[htpb]
    \centering
    \setlength{\tabcolsep}{1mm}
    \begin{tabular}{c|c|c|c|c|c|c|c|c}
        \hline
        \multirow{2}{*}{c/o} & \multicolumn{4}{c|}{Lesion-level}&\multicolumn{2}{c|}{Voxel-level}&\multicolumn{2}{c}{Dice-est./Actual} \\
        \cline{2-9}
         {}&{$prec.^{Pred}$}&{$rec.^{GT}$}&{$prec.^{IoU}$}&{$rec.^{IoU}$}&{$\mathcal{S}^{DSC}$}&{$\mathcal{D}^{HD}$}&{$\hat{FP}/FP$}&{$\hat{FN}$/FN}  \\
         \hline
         {0.1}&{0.26}&{0.69}&{0.12}&{0.17}&{0.33(0.18)}&{20.40(10.41)}&{257/409}&{86/106}\\
         \hline
         {0.2}&{0.36}&{0.61}&{0.21}&{0.27}&{0.34(0.19)}&{18.20(10.20)}&{214/306}&{101/136}\\
         \hline
         {0.3}&{0.42}&{0.54}&{0.23}&{0.27}&{0.34(0.20)}&{18.50(9.75)}&{232/273}&{103/154}\\
         \hline
         {0.4}&{0.44}&{0.52}&{0.25}&{0.29}&{0.33(0.20)}&{18.25(10.13)}&{256/253}&{111/169}\\
         \hline
         {0.5}&{0.47}&{0.49}&{0.25}&{0.28}&{0.33(0.21)}&{19.02(10.70)}&{216/237}&{104/178}\\
         \hline
         {0.6}&{0.49}&{0.45}&{0.25}&{0.28}&{0.32(0.21)}&{18.97(10.63)}&{219/213}&{101/194}\\
         \hline
         {0.7}&{0.52}&{0.42}&{0.27}&{0.27}&{0.31(0.21)}&{19.71(11.57)}&{179/190}&{101/208}\\
         \hline
         {0.8}&{0.58}&{0.37}&{0.26}&{0.26}&{0.29(0.21)}&{20.00(12.54)}&{132/158}&{98/227}\\
         \hline
         {0.9}&{0.64}&{0.31}&{0.27}&{0.23}&{0.26(0.21)}&{22.63(13.37)}&{114/115}&{101/255}\\
         \hline
       
    \end{tabular}
    \caption{Summary of the voxel- and lesion-level results, with varying cut-off (c/o) values. Estimated $\hat{FP}$ and $\hat {FN}$ are derived from linear relationship between DSC and lesion-level metrics. Results are based on 357 lesions in holdout patients.}
    \label{tab:summary}
\end{table*}

\section{Results}
\label{sec:results}
\begin{figure*}[bt]
    \centering
    \includegraphics[width=1\textwidth]{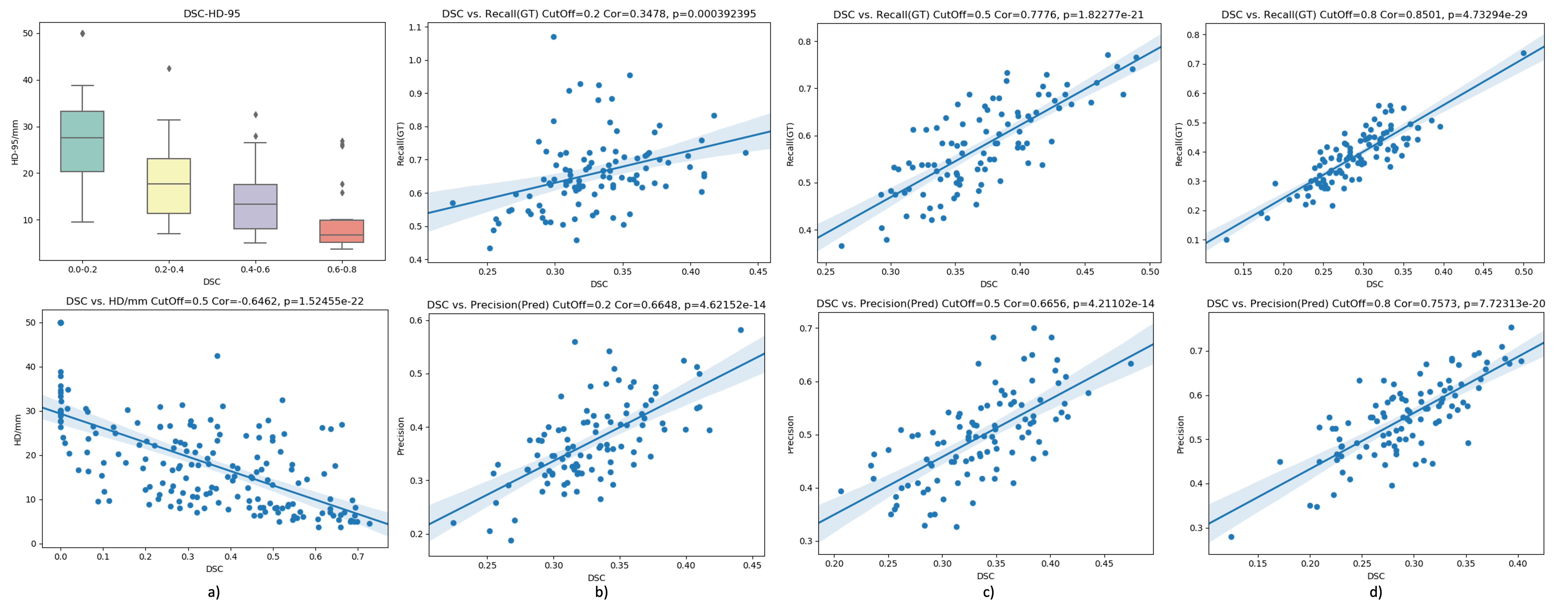}
    \caption{Quantitative comparison on holdout set, a) shows the correlation between $\mathcal{S}^{DSC}$ and $\mathcal{D}^{HD}$ in box plot and scatter plot, respectively. b-d) show the correlation between DSC and Precision/Recall under different cut-offs.}
    \label{fig:dicevsPR}
\end{figure*}

The segmentation results are reported in Table \ref{tab:summary}, showing that the $precision^{Pred}$ and $recall^{GT}$ are more sensitive to the cut-off than $precision^{IoU}$ and $recall^{IoU}$ are. This was indeed due to the difference between the symmetric overlap measure $\mathcal{S}^{IoU}$/$\mathcal{S}^{DSC}$ and the asymmetric $\mathcal{S}^{GT}$/$\mathcal{S}^{Pred}$, discussed in Secs.~\ref{sec:metrics_vox_lesion} and~\ref{sec:metrics_asymmetric}. Fig.\ref{fig:dicevsPR} provides examples of the lesion segmentation with their quantitative results.

\subsection{Comparison between DSC and HD}
Fig.\ref{fig:dicevsPR} a) observe a monotonic relationship between $\mathcal{S}^{DSC}$ and $\mathcal{D}^{HD}$, with a correlation of $r$=-0.6462 ($p-value$=0.000) and a ``moderate'' pairwise agreement of $\kappa$=0.57. However, examples were found that the two metrics disagree on which case of a given pair is a ``better" segmentation. For example, the case in Fig.\ref{fig:cat} c) yielded a lower $\mathcal{S}^{DSC}$ and a lower $\mathcal{D}^{HD}$, compared with those measured from the cases in a) and b). This indeed indicates that variable number of ROIs may be a factor of such disagreement, due to their sizes and locations.

\subsection{Comparison between voxel- and lesion-level metrics}
Fig.\ref{fig:dicevsPR} b-d) illustrate that both lesion-level metrics are increasingly correlated with $\mathcal{S}^{DSC}$ with higher cut-offs, ranging from 0.35 to 0.85 and from 0.66 to 0.76, for $recall^{GT}$ and $precision^{Pred}$, respectively.

\begin{figure*}[t]
    \centering
    \includegraphics[width=1\textwidth]{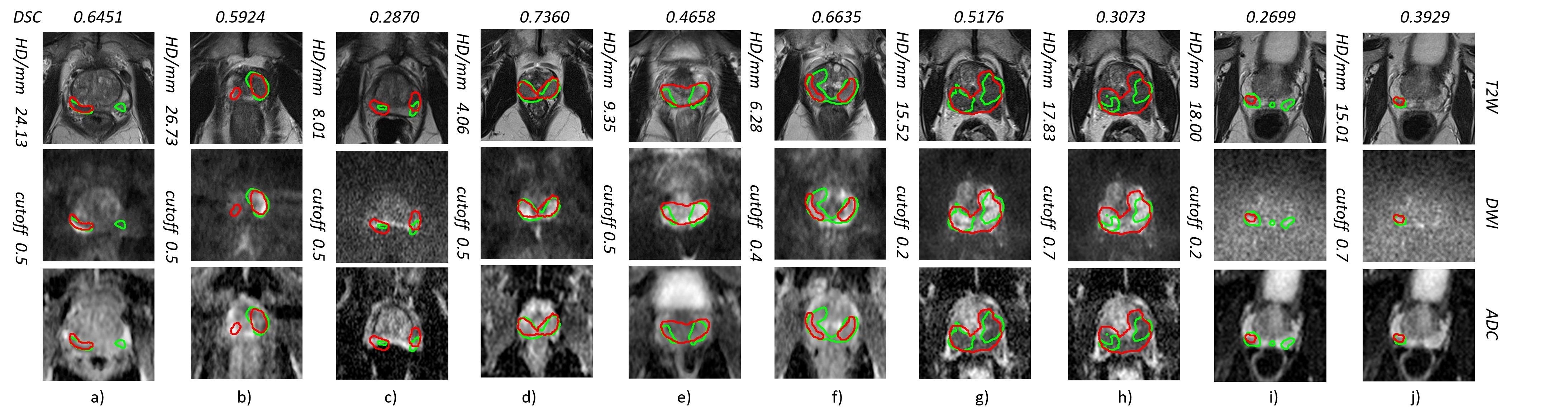}
    \caption{Holdout set segmentation results, with ground-truth (predicted) lesions in red (green), showing examples of a-c) inconsistency between DSC and HD, d) good prediction in all evaluation metrics, e) single lesion with multiple predictions, f) two lesions with one prediction, g-h) poorer performance with higher cut-off, and i-j) better performance with higher cut-off.}
    \label{fig:cat}
\end{figure*}

In Fig.\ref{fig:cat}, Cases d-f) show examples of levels of correlation and agreement between the voxel- and lesion-level metrics, Cases g-j) show the effect of different cut-off values on two cases. It is noteworthy that these cases were cherry-picked to show various scenarios that motivated $recall^{GT}$/$precision^{Pred}$ and the overall comparison reported in Fig\ref{fig:dicevsPR}. Visual examples are generally consistent with the reported quantitative correlation and agreement results and showed an interestingly strong correlation between the voxel- and lesion-level evaluation metrics.

As summarised in the last two columns of Table \ref{tab:summary}, the ratio of Dice-estimated false-positives to true false-positives rises, partly due to the reduced number of false-negative lesions as the cut-off increases. The DSC may therefore be considered a good evaluation metric at a cutoff close to 0.9. The opposite trend can be observed for the false-negatives. However, the differences in general cannot be overlooked, with the maximum being 37\% (152) and 60\% (154) of the respective true errors.

\section{Conclusion}
In this work, we compare voxel-level segmentation metrics DSC and HD, with lesion-level accuracy metrics, on the prostate lesion segmentation task. For evaluating segmentation output, we proposed new lesion detection metrics that are asymmetric and suitable for voxel-level cut-off adjustment. Experimental results show considerable correlation and pairwise agreement not only between the DSC and HD, but also between voxel- and lesion-level metrics. Notwithstanding a degree of agreement and the apparent correlations, the voxel-level segmentation metrics could still lead to significant misinterpretation in lieu of the lesion-level errors. Following the presented evidence from the real-world clinical application, we recommend reporting these voxel-level metrics with caution and an appreciation of their limitations. Future work includes studies with wider clinical downstream tasks that use automated segmentation and a comparison between these metrics as loss functions for network training.

\section*{Acknowledgment}
This work was supported by the International Alliance for Cancer Early Detection, an alliance between Cancer Research UK [C28070/A30912; C73666/A31378], Canary Center at Stanford University, the University of Cambridge, OHSU Knight Cancer Institute, University College London and the University of Manchester. This work was also supported by the Wellcome/EPSRC Centre for Interventional and Surgical Sciences [203145Z/16/Z].

%
%
%
\bibliographystyle{splncs04}
\bibliography{myref}

\end{document}